\documentclass{raa_twocolumn}
\usepackage{graphicx,times}             
\graphicspath{{./figures/}}
\usepackage{natbib}
\usepackage{amssymb,amsmath}
\bibpunct{(}{)}{;}{a}{}{,}
\usepackage[pagebackref=true]{hyperref}
\usepackage{comment}
\usepackage{xspace}
\usepackage{xcolor}
\definecolor{myred}{HTML}{C41010}
\definecolor{mycyan}{HTML}{228888}
\definecolor{mygreen}{HTML}{558822}
\definecolor{mybrown}{HTML}{884400}
\definecolor{myblue}{HTML}{00007f}

\newcommand{\fpm}[1]{\textcolor{black}{#1}}
\newcommand{\eg}{\textit{eclgrm}\xspace}
\newcommand{\egv}{\textit{eclgrm-vhf}\xspace}
\newcommand{\egx}{\textit{eclgrm-xband}\xspace}
\newcommand{\egui}{\textit{eclgrm-ui}\xspace}
\newcommand{\egis}{ECLGRM-IS\xspace}

\begin{document}

\title{The GRB joint scientific analysis pipeline of the ECLAIRs and GRM instruments on board SVOM}

   \volnopage{Vol.0 (202x) No.0, 000--000}      
   \setcounter{page}{1}          

   \author{F. Piron 
     \inst{1,*}\footnotetext{$*$Corresponding Authors, these authors contributed equally to this work.}
     \and F. Daigne
     \inst{2,3,*}
     \and T. Maiolino
     \inst{1}
     \and P. Maeght
     \inst{1}
     \and U. Jacob
     \inst{1}
     \and M.G. Bernardini
     \inst{1,4}
     \and D. Corre
     \inst{2}
     \and J. Wang
     \inst{2}
     \and F. Lacreu
     \inst{2}
     \and G. Tcherniatinsky
     \inst{2}
     \and L. Domisse
     \inst{2}
     \and T. Barlyaeva
     \inst{1}
     \and A. Maïolo
     \inst{1}
     \and J.-L. Atteia
     \inst{5}
     \and L. Bouchet
     \inst{5}
     \and M. Brunet
     \inst{5}
     \and J.-P. Dezalay
     \inst{5}
     \and O. Godet
     \inst{5}
     \and S. Guillot
     \inst{5}
     \and H. Yang
     \inst{5}
     \and B. Arcier
     \inst{5}
     \and S. Mate
     \inst{5}
     \and N. Dagoneau
     \inst{6}
     \and L. Jouvin
     \inst{6}
     \and K. Tazhenova
     \inst{6}
     \and T. Sadibekova
     \inst{7}
     \and P. Bacon
     \inst{8}
     \and N. Bellemont
     \inst{8}
     \and F. Cangemi
     \inst{8}
     \and A. Coleiro
     \inst{8}
     \and J. He
     \inst{9}   
     \and Y. Huang
     \inst{9}   
     \and L. Li
     \inst{9}   
     \and H. Shi
     \inst{9}
     \and J. Wang
     \inst{9}   
     \and P. Wang
     \inst{9}   
     \and L. Zhang
     \inst{9}   
     \and X.-Y. Zhao
     \inst{9}   
     \and S. Zheng
     \inst{9}   
}
   \institute{Laboratoire Univers et Particules de Montpellier, Université Montpellier, CNRS/IN2P3, F-34095 Montpellier, France; {\it piron@in2p3.fr}\\
    \and
    Sorbonne Université, CNRS, UMR 7095, Institut d’Astrophysique de Paris, 98 bis bd Arago, 75014 Paris, France; {\it daigne@iap.fr}\\
    \and
    \fpm{Institut Universitaire de France;}\\
    \and
    \fpm{INAF --Osservatorio Astronomico di Brera, Via Bianchi 46, I23807 Merate, LC, Italy;}\\
    \and
    Institut de Recherche en Astrophysique et Planétologie, Université de Toulouse/CNRS/CNES, 9 avenue du colonel Roche 31028 Toulouse, France;\\
    \and
    CEA Paris-Saclay, Institut de Recherche sur les lois Fondamentales de l’Univers, 91191 Gif sur Yvette, France;\\
    \and
    Université Paris-Saclay, Université Paris Cité, CEA, CNRS, AIM, 91191 Gif-sur-Yvette, France;\\
    \and
    Université Paris Cité, CNRS, CEA, Astroparticule et Cosmologie, F-75013 Paris, France;\\
    \and
    State Key Laboratory of Particle Astrophysics, Institute of High Energy Physics, Chinese Academy of Sciences, Beijing 100049, China\\
\vs\no
   {\small Received 202x month day; accepted 202x month day}}

   \abstract{
       The study of the prompt high-energy emission of Gamma-Ray Bursts (GRBs) with SVOM relies on the observations
       performed by ECLAIRs (4--150\,keV) and the Gamma-Ray Monitor (GRM, 0.015--5\,MeV), the two wide field-of-view
       instruments on board the satellite. 
       In this article, we introduce the \eg pipelines running at the French Science Center of SVOM, which combine the
       ECLAIRs and GRM data to generate scientific \fpm{data} products describing the GRB broad-band temporal and spectral properties. 
       The architecture of the pipelines is presented, as well as their activation following each onboard trigger, and
       their workflow.
       The statistical data analysis methods employed by the pipelines are described, along with the 
       scientific \fpm{data} products that are created in real time or from the full event data.
       We also present the \egui user interface which allows the scientists on shift to monitor the automated data
       processings in the pipelines, and to optimize the analysis results interactively.
}

   \authorrunning{F. Piron, F. Daigne \& al.}            
   \titlerunning{The SVOM ECLGRM GRB analysis pipeline}  

   \maketitle

%
%


\section{Introduction}
\label{sect:intro}

The SVOM (Space-based multi-band astronomical Variable Objects Monitor) observatory \citep{raa-mission-space,
  raa-mission-ground, raa-mission-satellite} is primarily dedicated to Gamma-Ray Burst (GRB)
science \citep{white-paper}.
The two wide field-of-view instruments on board the satellite, the coded-mask camera ECLAIRs (4--150\,keV,
\cite{raa-ecl-inst}) and the Gamma-Ray Monitor (GRM, 0.015--5\,MeV, \cite{raa-grm-inst}),
are monitoring the transient sky and detecting GRBs in X-ray and gamma-ray energy bands, allowing to characterize the
temporal and spectral properties of the GRB high-energy prompt emission.
The satellite of SVOM has been placed on a low-Earth orbit, with a quasi-anti solar pointing to favor rapid follow-up
observations of SVOM GRBs by ground telescopes. This implies periodic passages of the Earth in the
fields of view of the instruments on board SVOM \citep{raa-mission-space}.

The \fpm{ECLAIRs-GRM} system represents a step forward in the spectral characterization of the population
of long and short GRBs, allowing the investigation of a possible thermal component superimposed to
their non-thermal emission \citep{Bernardini+2017} or of spectral breaks distinctive of synchrotron emission
(e.g. \citealt{Oganesyan+2017}).
The search for these additional components will benefit mainly from the ECLAIRs low-energy detection threshold at 4\,keV,
but also from the combination of ECLAIRs and GRM data covering a broad energy range.
More than one hundred GRBs are detected each year by ECLAIRs and/or the GRM, which will also provide new insights
on the GRB class, as for example the spectral characterization of the short GRB sub-class showing a soft extended emission
after the initial hard spike \citep{Bernardini+2017}.
The first GRB detections described in \cite{raa-grbs} already demonstrate the ability to characterize the GRB population
in all its diversity.

\begin{figure}[t!]
  \centering
  \includegraphics[width=\linewidth]{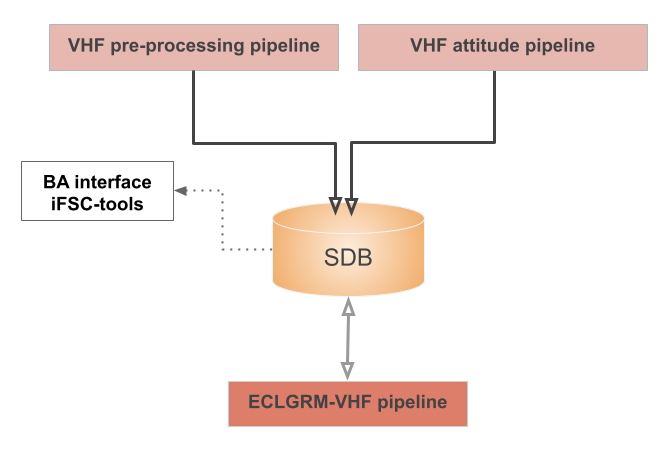}
  \caption{
    \label{fig:eclgrm-vhf}
    Functional diagram of the \egv pipeline.
    The two FSC pipelines in pink color generate \fpm{SDP}s that are retrieved and used by the \egv pipeline
    through the FSC SDB. 
  }
\end{figure}

In order to \fpm{best} exploit the potential of combining ECLAIRs and GRM observations of GRBs, two \fpm{pipelines (\egv 
  and \egx) have been developed to analyze their data immediately once they are available at the SVOM French Science
  Center (FSC) \citep{raa-fsc}. 
The pipelines have been deployed in Docker containers running at the FSC, and they use} a common library of algorithms to
create the \fpm{Scientific Data Products (SDPs hereafter,} see \cite{raa-fsc, raa-fsc-tools}) characterizing the GRB prompt
high-energy emission. 
A user interface (\egui) has also been developed to monitor the data processings in the pipelines and, if needed, to
restart some processes manually to refine the results:
\begin{itemize}
\item The GRB quick analysis pipeline (\egv) processes the ECLAIRs and/or GRM count rate light curves that are built on
  board following each GRB alert and quickly transmitted to the ground via VHF antennas \citep{raa-vhf-system}.
  For each instrument, and also for the joint ECLAIRs and GRM analysis, the pipeline creates the related quick
    \fpm{SDP}s, which are used to promptly characterize the source and help validate the trigger as a genuine GRB.
\item The GRB complete analysis pipeline (\egx) processes the ECLAIRs and GRM event data which are registered on board and
  transmitted to the ground via X-band antennas several times per day \citep{raa-ground-system}.
  The pipeline performs a complete temporal and spectral analysis of the GRB, for ECLAIRs alone and for the
  \fpm{ECLAIRs-GRM} system, and it creates the related X-band \fpm{SDP}s.
  The X-band \fpm{SDP}s that are based on GRM data only are computed by the GRM pipeline running at the GRM
    Instrument Center (GRM-IC) in China \citep{raa-grm-pipeline}, using a similar methodology and a common \fpm{SDP} data model
    \citep{raa-fsc}.
\item The \egui user interface is used by the ECLAIRs-GRM Instrument Scientist (\egis hereafter) on shift, who
  is an expert of the software for ECLAIRs and GRM data analysis.
  The role of the IS is to monitor the automated processings in the pipelines, to inspect the analysis results and improve
  them if needed, and to validate the output \fpm{SDP}s \citep{raa-fsc-tools}.
\end{itemize}

\begin{figure}[t!]
  \centering
  \includegraphics[width=\linewidth]{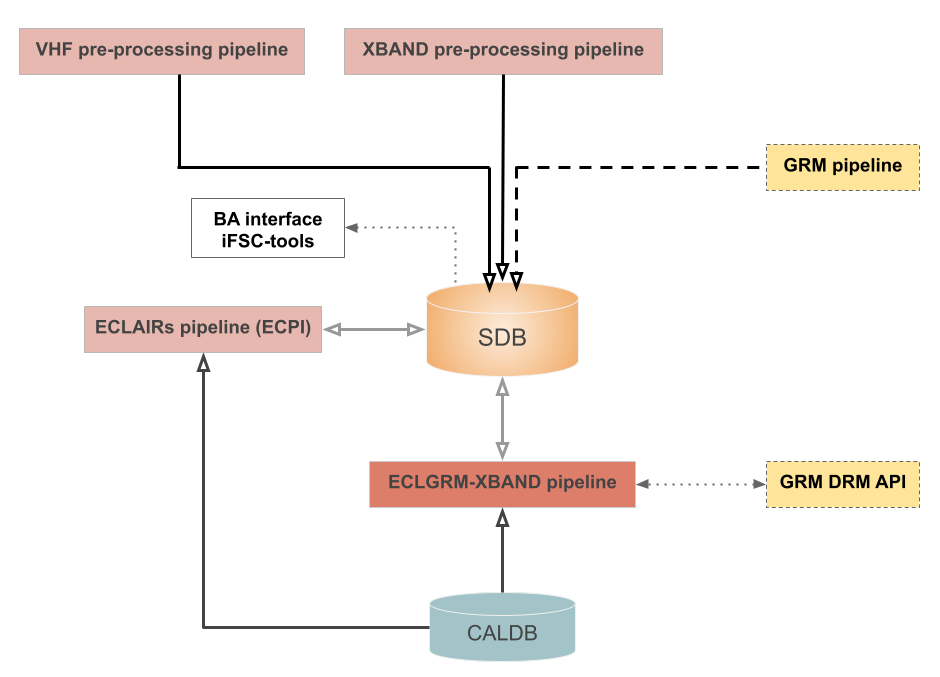}
  \caption{
    \label{fig:eclgrm-xband}
    Functional diagram of the \egx pipeline.
    The three FSC pipelines in pink color create \fpm{SDP}s that are retrieved and used by the \egx pipeline through the FSC SDB.
    The two services running at the GRM-IC \citep{raa-grm-pipeline, raa-grm-rsp} are indicated in yellow color: the GRM
    pipeline creates the GRM calibrated event data products, which are transferred to the FSC SDB, while
    requests to the GRM DRM API are sent directly. 
  }
\end{figure}

In section~\ref{sect:archi} we describe the architecture of the pipelines. The quick and complete analysis pipelines are
described in sections~\ref{sect:vhf} and~\ref{sect:xband}, respectively. In section~\ref{sect:ui}, we summarize the
\fpm{functionalities of the user interface} and their use by the \egis.
Section~\ref{sect:conclusion} presents our conclusions and the possible future pipeline improvements.
\fpm{A table listing all acronyms is given in Appendix~\ref{sect:appendix}.}

\section{Pipeline architecture and activation}
\label{sect:archi}
\begin{table*}[t!]
  \begin{center}
    \caption{
      \label{tab:eclgrm-vhf}
      Time Resolutions and Energy Bands in the \egv Pipeline.
      The mid and best time resolutions for the OBLCs are available only in a limited time interval around the trigger
      time.
      The OBLC time resolutions and energy bands are pre-defined on board.
      The QT90 and QHR energy bands are the default values used in the pipeline automated mode.
    }
    \begin{tabular}{lccc}
      \hline\noalign{\smallskip}
      \fpm{Time resolutions and energy bands}& VHF-ECL target& VHF-GRM target\\
      \hline\noalign{\smallskip}
      OBLC pre-defined time resolutions (s) & low: 6.4 ; mid: 0.8 ; best: 0.1  & low: 6.4 ; best: 0.8  ; high$^{*}$: 0.001   \\
      OBLC pre-defined energy bands (keV)  & 1: 5-8 ; 2: 8-20 ; 3: 20-50 ; 4: 50-120      & 1: 4-550 ; 2: 550-5000    \\
      \hline\noalign{\smallskip}
      QT90 default energy band               & 1+2+3+4   & 1    \\
      \hline\noalign{\smallskip}
      QHR default energy bands                 & 4/1 ; 4/2 ; 4/(1+2) & 2/1  \\
      \hline\noalign{\smallskip}
      \hline\noalign{\smallskip}
      & \multicolumn{2}{c}{VHF-ECLGRM target}  \\
      \hline\noalign{\smallskip}
      QHR default energy bands & \multicolumn{2}{c}{(GRM 2)/(ECL 1+2+3+4), (GRM 1)/(ECL 1), (GRM 1)/(ECL 1+2)}  \\
      \hline\noalign{\smallskip}
    \end{tabular}
  \end{center}
  $^{*}$ the high time resolution GRM OBLC covers an interval of $\pm 2$ s around the trigger time, and is built from the
  sum of the three GRD count rates. This light curve is rebinned with a 50\,ms time resolution that is adapted to GRB
  durations smaller than 0.8\,s.
\end{table*}

The onboard trigger and observation scenarios (ECLAIRs and/or GRM trigger, satellite autonomous slew
to place bright GRB candidates detected onboard by ECLAIRs at the center of its field of view) result in different 
types of VHF and X-band data received on the ground and pre-processed upstream of the \eg pipelines.
Depending on which input \fpm{SDP}s are available at a given time,
the FSC orchestrator \citep{raa-fsc} activates specific \eg processes, referred to as `targets' hereafter.
The \egv pipeline features a total of three targets, while the \egx pipeline features two targets.
Besides this automated processing mode that is activated after each onboard trigger, the pipelines can also be restarted
manually through the \egui with different parameters.

Figures~\ref{fig:eclgrm-vhf} and ~\ref{fig:eclgrm-xband} show the functional diagrams of the \egv and \egx pipelines, respectively.
The pipelines need products from the FSC VHF or X-band pre-processing pipelines \citep{raa-fsc}, which convert the
  instrument and satellite raw data into \fpm{SDP}s in FITS format.
Two other FSC services are used: the Science Database (SDB) to fetch or send \fpm{SDP}s, and the
  Calibration Database (CALDB) to retrieve the ECLAIRs response files \fpm{that are needed by the spectral
    analysis in the \egx pipeline}.
In addition, the \egx pipeline is fed with the ECLAIRs calibrated event data that are produced by the quick-look
  automated analysis of the \fpm{ECLAIRS pipeline (ECPI, \cite{raa-ecpi})}, and it has two external dependencies at the GRM-IC: (i) the GRM
  calibrated event data that are produced by the GRM pipeline and then transferred to
the FSC SDB via the Chinese Science Center of SVOM \citep{raa-ground-system},
and (ii) an API to retrieve the response files for the three Gamma-Ray Detectors (GRDs) of the GRM
  \citep{raa-grm-inst, raa-grm-rsp}. 
  
Each target follows a standardized workflow consisting in a series of successive analysis tasks, each of them being
designed to produce a specific set of \fpm{SDP}s. Each task operates as follows: retrieval of the \fpm{SDP} FITS templates,
download of the input files, initialization of the task parameters, execution of the scientific algorithms, filling of the
output \fpm{SDP} FITS files \fpm{with the analysis results (computed physical quantities, see the next sections)}, export
to the SDB along with the ancillary files.
The latter include figures and control plots, which are generated at each step of the data analysis sequence to
assess the quality of the results.
\fpm{More detail on the SDP level definition can be found in \cite{raa-ground-system} and \cite{raa-fsc-tools}, while
  \cite{raa-fsc} presents the SDP data model and the use of their FITS templates.}


\section{The GRB quick analysis pipeline}
\label{sect:vhf}
\subsection{\fpm{Purpose}}

The \egv pipeline uses the ECLAIRs and GRM data received via the VHF network to compute rapidly all the necessary \fpm{SDP}s 
to characterize the prompt high-energy emission of a GRB candidate in real time: GRB and background
count light curves, emission duration, on-ground detection significance, peak fluxes, hardness
ratios and early classification (likely GRB type or other astrophysical event). 
\fpm{This scientific analysis is very rapid and takes less than one minute, typically.}

\subsection{\fpm{Activation Targets and Dependencies}}
The three pipeline targets are VHF-ECL and VHF-GRM, for processing ECLAIRs and GRM onboard light
curves, respectively, and VHF-ECLGRM for the joint analysis.
The input \fpm{SDP}s of the first two targets are created at the FSC (see Figure~\ref{fig:eclgrm-vhf}):
\begin{itemize}
\item The VHF pre-processing pipeline provides the time and localization of the onboard
  alert(s), as well as the onboard count light curves (OBLCs).
  The OBLCs are built around the trigger time in different pre-defined energy bands and with pre-defined time resolutions
  (see Table~\ref{tab:eclgrm-vhf}).
\item The VHF attitude pipeline provides the \fpm{SDP}s containing the satellite attitude and orbital position, which are unevenly
  sampled around the trigger time. 
\end{itemize}

The number of activated targets, and the corresponding data flow and \fpm{SDP} outputs, depend on the
SVOM trigger scenario \citep{raa-ecl-trigger, raa-grm-trigger}:
\begin{itemize}
\item The VHF-ECL target is activated when ECLAIRs OBLCs are available in the FSC SDB, which happens only if ECLAIRs
  triggered over a timescale shorter than 10\,min.
\item The VHF-GRM target is always activated since GRM OBLCs are built in all trigger scenarios. In the specific case
  where ECLAIRs triggers first, the GRM trigger is inhibited (to download only the best alert localization to the ground),
  and the possible GRM detection of the GRB candidate is assessed by the \egv pipeline.
\end{itemize}

The successive tasks executed by the \egv pipeline for the VHF-ECL and VHF-GRM targets are described below. For the
VHF-ECLGRM target, only the hardness ratio and early classification tasks are performed.
\begin{figure}[t!]
  \centering
  \includegraphics[width=\linewidth]{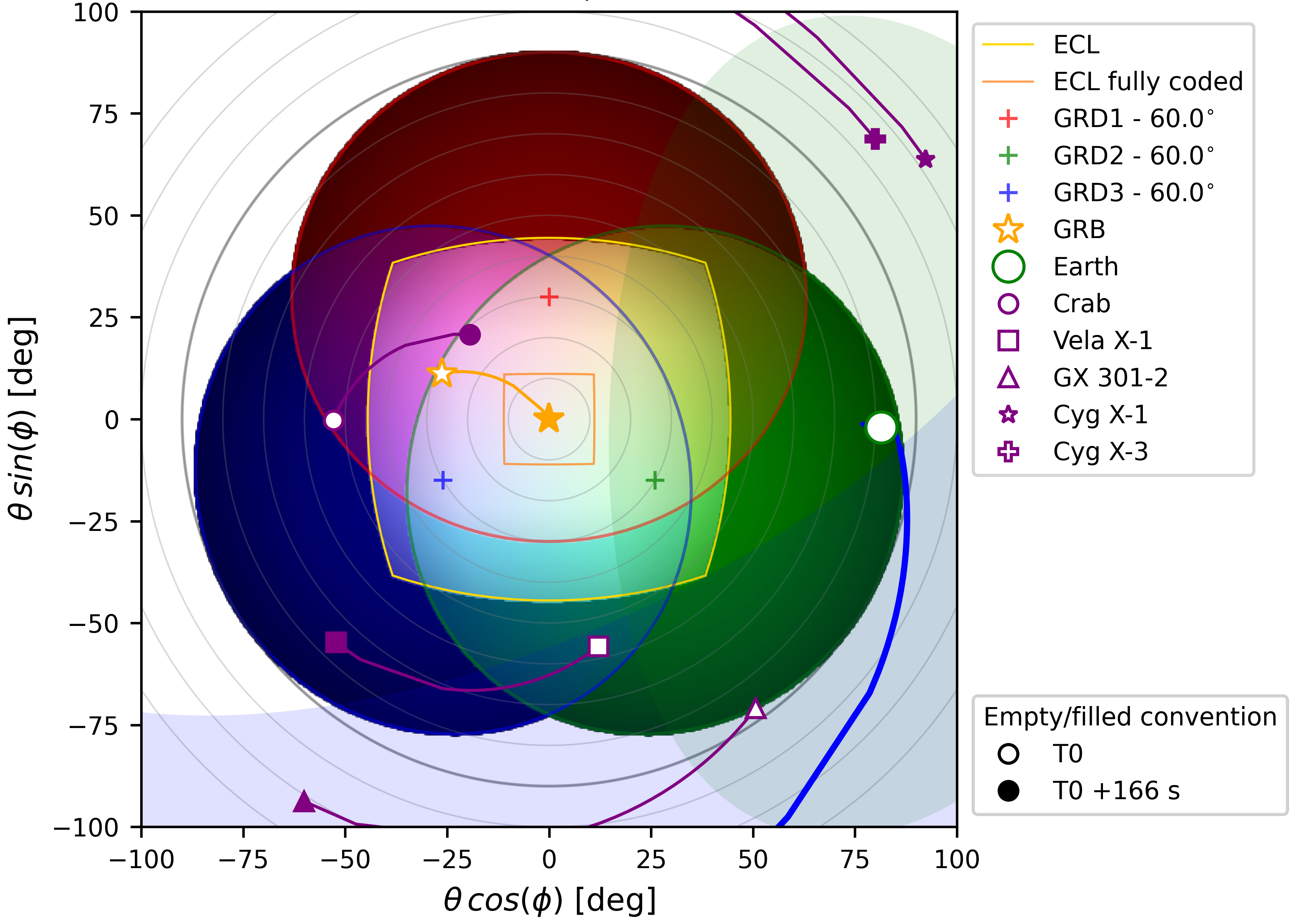}
  \caption{
    \label{fig:250205a_fov}
     GRB\,250205A (burst-id sb25020504) in the ECLAIRs and GRM fields of view (FoVs):
     localization in the partially- and fully-coded FoVs of ECLAIRs, and in the GRD FoVs (with a radius arbitrarily set to
     60$^\circ$).
    The GRB trajectory is displayed from the trigger time T$_0$ (empty orange star) to the end of the slew (filled orange
    star), as well as bright known X-ray sources.
    The sky portions occulted by the Earth at theses times are displayed by the light green and blue shaded areas, respectively.
  }
\end{figure}
\begin{figure}[t!]
  \centering
  \includegraphics[width=\linewidth]{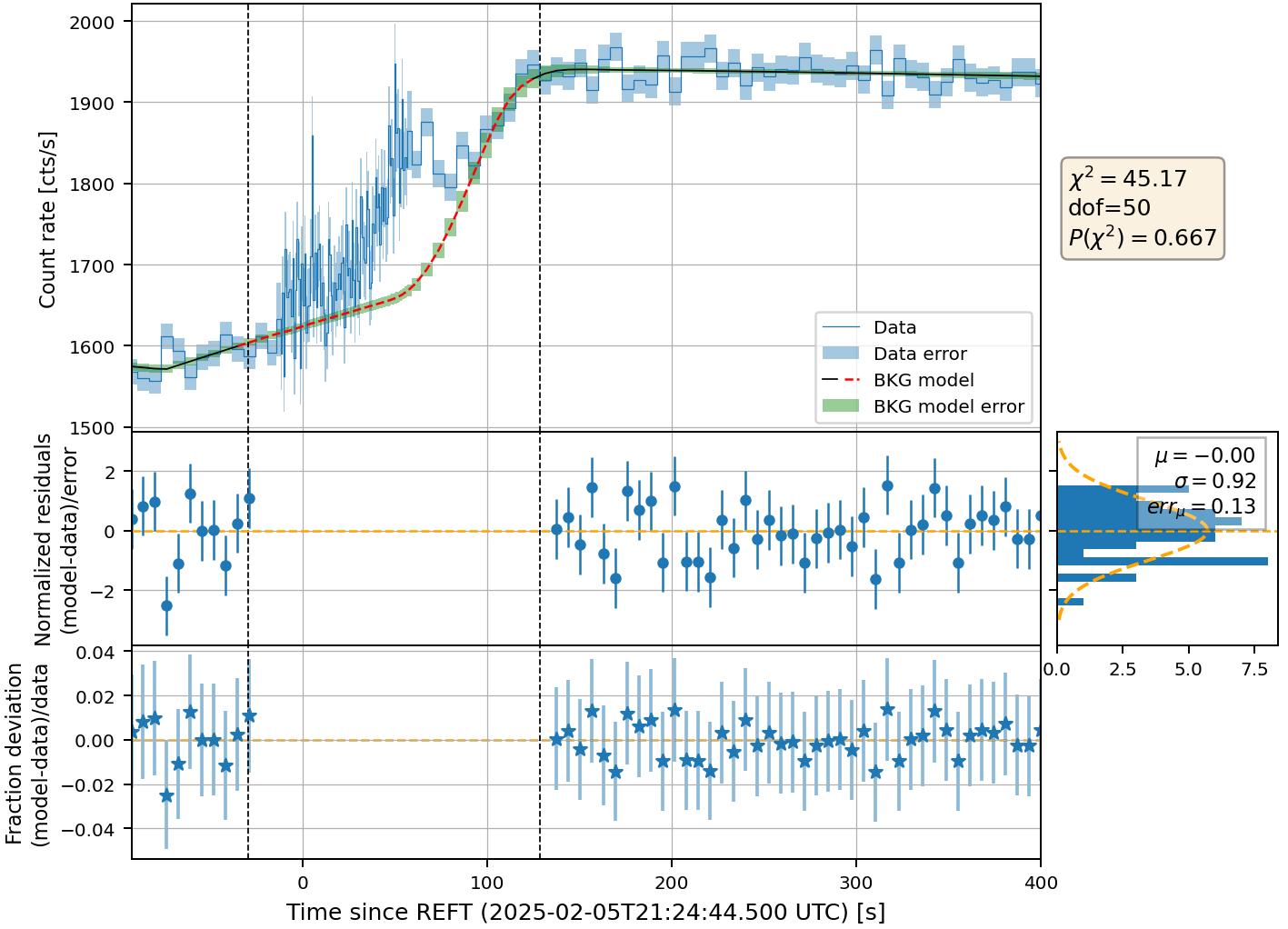}
  \caption{
    \label{fig:250205a_bkg_fit}
    GRB\,250205A (burst-id sb25020504) automated analysis by the \egv pipeline: the top panel shows the count 
    light curve (CLC) in the GRM/GRD1 4--550\,keV energy band. The two background intervals (left and right) are
    indicated, as well as the best-fit background Model-E.
    The mid and bottom panels show the normalized residuals of the fit and the deviation between the background data and
    the best-fit model, respectively.
  }
\end{figure}
\begin{figure*}[t!]
  \centering
  \includegraphics[width=0.8\linewidth]{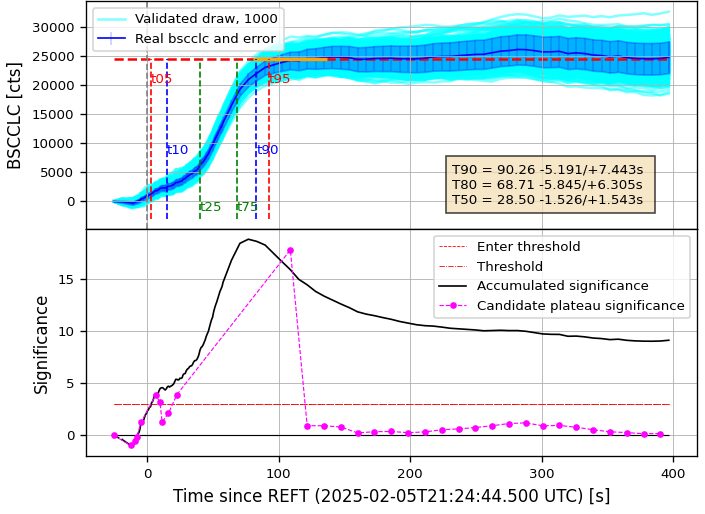}
  \caption{
    \label{fig:250205a_resume_vhf}
     GRB\,250205A (burst-id sb25020504) automated analysis by the \egv pipeline: the top panel shows the
     background-subtracted cumulative count light curve (BSCCLC, in blue) in the GRM 4--550\,keV energy band.
     The 1000 BSCCLCs simulated through resampling are shown in cyan.
     The horizontal dashed red line indicates the GRB count fluence.
     The bottom panel shows the significance of the accumulated signal between the current plateau and a new plateau 
       candidate (magenta filled circles), and from the start of the light curve (black line). 
  }
\end{figure*}

\subsection{\fpm{Temporal Analysis Tasks}}
The Quick Light Curve (QLC) task first creates informative plots to help understand the observational context
around the trigger time: summary of ECLAIRs onboard alert(s), pointing directions of the instruments, angular
positions of the GRB and of nearby known X-ray sources in the fields of view of ECLAIRs and of the three GRDs (see 
Figure~\ref{fig:250205a_fov}), GRB angular distance to the Earth's limb, and possible source occultation by the
  Earth.
Then, it creates count light curves (CLCs) in different energy
bands from the OBLCs, using the best (smallest) time resolution available at any time.
By using the Bayesian Blocks method \citep{Scargle+2013}, it identifies the two optimal background time intervals
  (before and after the GRB signal) for each CLC, along with the best-fit background model, and it calculates the
background-subtracted count light curves (BSCLCs).
Currently, the pipeline considers two background models:
Model-T, a polynomial function of time, and Model-E, a polynomial function of the cosine of the angle $\theta_E(t)$
between the detector's main axis and the direction of Earth center.
Both models have been tested using GEANT4 simulations of the main sources of background in orbit: the cosmic X-ray
background (CXB), its reflection on the Earth's atmosphere, and the Earth's albedo \citep{Mate+2019,Maiolo-phd}.
As the SVOM satellite follows a quasi-antisolar pointing, the sky fraction occulted by the Earth varies along the orbit
and modulates the background rate (see \cite{raa-ecl-inst}).
These variations can be more rapid in case of slew, where Model-T often fails while Model-E provides good background fits,
as shown in Figure~\ref{fig:250205a_bkg_fit}. 

The Quick T90 duration (QT90) task estimates the GRB signal duration in a selected energy band (see Table~\ref{tab:eclgrm-vhf}).
Following the QLC task, it simulates one thousand CLCs through Poisson resampling, and background-subtracted cumulative
count light curves (BSCCLCs).
The plateau at the end of each BSCCLC is identified automatically (see Figure~\ref{fig:250205a_resume_vhf}), allowing to
compute the distribution of the times $t_x$ at which $x\%$ of the total signal counts is reached, and the distribution of
the durations $QT_\mathrm{90}=t_{95}-t_{05}$, $QT_\mathrm{80}=t_{90}-t_{10}$, and $QT_\mathrm{50}=t_{75}-t_{25}$, which contain 90\%, 80\%,
and 50\% of the GRB counts, respectively.
The final times and durations are given as the median of their respective distributions, with $68\%$ confidence intervals. 
The QT90 task also calculates the on-ground significance of the GRB signal in the $QT_\mathrm{90}$ time interval
following the prescription of \cite{Vianello2018}. 

\subsection{\fpm{Final Tasks}}

The Quick Peak Flux (QPF) task computes, for each BSCLC generated by the QLC task, the maximum count rate
(in counts/s) along with its respective error, and the time at which this peak is observed.

The Quick Hardness Ratio (QHR) task computes the ratio of the total numbers of signal counts
in the $QT_\mathrm{90}$ time interval between two energy bands (high/low) as defined in Table~\ref{tab:eclgrm-vhf}.

The Crude Classification (CRCLASS) task is being developed to classify ECLAIRs and GRM triggers automatically 
(as long or short GRBs, magnetar flares, solar flares, particle events, etc.) through a Bayesian classification scheme
  based on the onboard information and VHF products from the previous tasks.

\section{The GRB complete analysis pipeline}
\label{sect:xband}

\subsection{\fpm{Purpose}}

The \egx pipeline uses the ECLAIRs and GRM complete event data to perform a \fpm{full} temporal and spectral study of the
GRB prompt high-energy emission, and to create the related X-band \fpm{SDP}s:
light curves in different energy bands, final estimates of the emission duration, time-integrated and time-resolved
reconstruction of the GRB photon spectra (assuming different models) and associated by-products (broad-band spectral
energy distribution, fluxes and fluences), time-dependent hardness ratios, and spectral lags.
It also computes the rest-frame luminosity and energetics if the source redshift has been measured.
\fpm{The analysis of ECLAIRs data takes a few minutes only, while the joint analysis of ECLAIRs and GRM data can be
  completed in 15 minutes, typically.}

\subsection{\fpm{Activation Targets and Dependencies}}

The two pipeline targets are XBAND-ECL to process ECLAIRs data, and XBAND-ECLGRM for a joint analysis with the GRM.
Their input \fpm{SDP}s are created at or outside the FSC (see Figure~\ref{fig:eclgrm-xband}):
\begin{itemize}
  \item The FSC VHF pre-processing pipeline provides the onboard alert information (see section~\ref{sect:vhf}).
  \item The FSC X-band pre-processing pipeline provides the ECLAIRs \fpm{non calibrated event data}, as well
    as the satellite attitude and orbital position sampled every second.
  \item The FSC ECPI pipeline \citep{raa-ecpi} provides the ECLAIRs calibrated event data (ECL-EVT-CAL).
    Moreover, the \egx container includes the ECPI sofware to call specific modules directly.
  \item The GRM pipeline \citep{raa-grm-pipeline} provides the calibrated event data of each GRD (GRM-EVT-TRIG).
  \item The FSC CALDB \citep{raa-fsc} provides the ECLAIRs response files (on-axis effective area, off-axis efficiency, energy
    redistribution).
  \item The GRM DRM API \citep{raa-grm-rsp} provides the \fpm{Detector Response Matrix (DRM) of each GRD}, which are
    computed for each GRB to account for the indirect signal induced by the scattering of the source photons on the
    Earth's atmosphere. 
    Unlike the previous services whose output products are retrieved from the FSC SDB, this API is called directly
    by the \egx pipeline (see Figure~\ref{fig:eclgrm-xband}). 
\end{itemize}

\begin{figure}[t!]
  \centering
  \includegraphics[width=\linewidth]{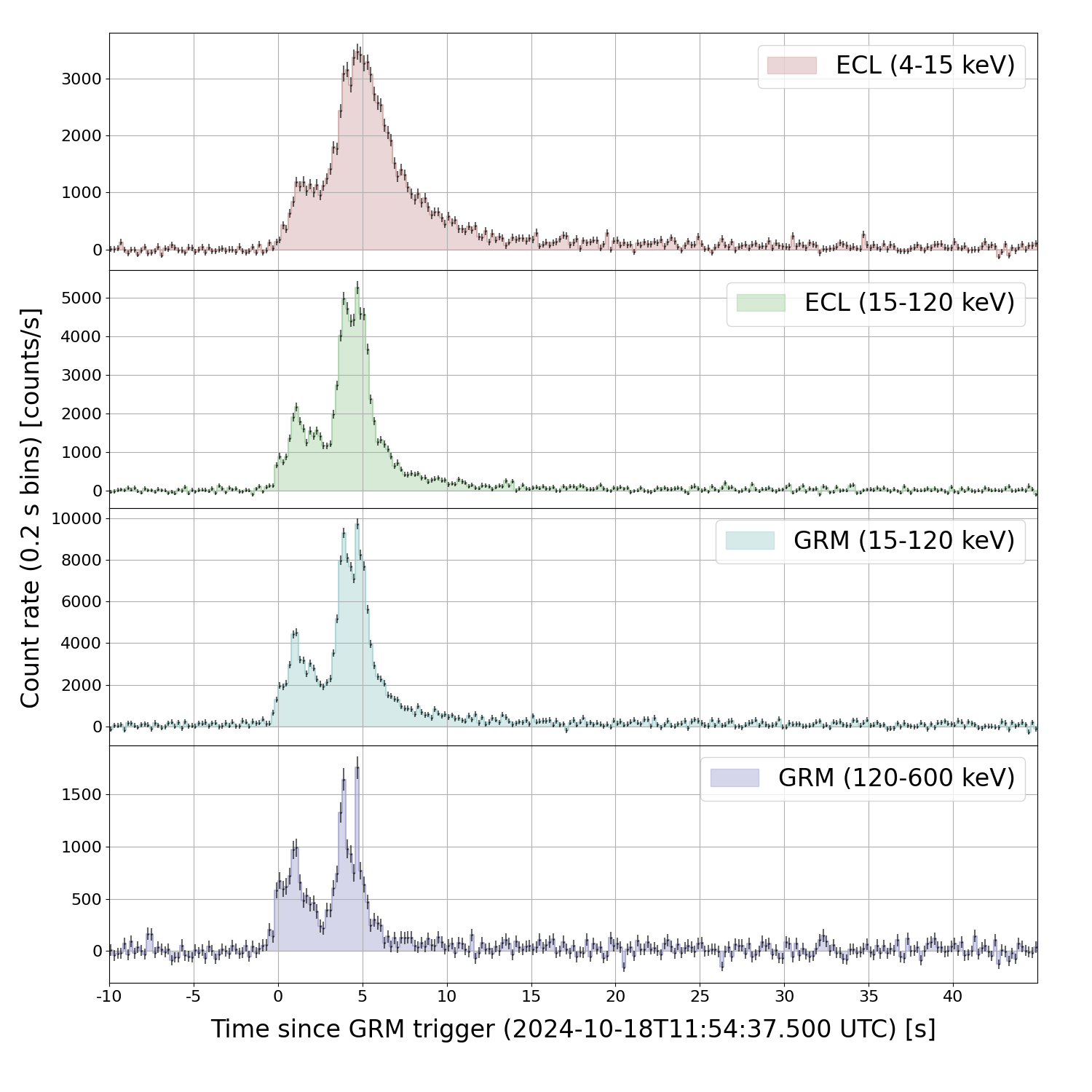}
  \caption{
    \label{fig:241018a-xband_eclgrm_clc}
    GRB\,241018A (burst-id sb24101802) analysis by the \egx pipeline:
    background-subtracted count light curves (BSCLCs) in ECLAIRs and the GRM in different energy bands.
    The ECLAIRs light curves are based on the detection plane pixels illuminated by the GRB.    
  }
\end{figure}
\begin{figure}[t!]
  \centering
  \includegraphics[width=\linewidth]{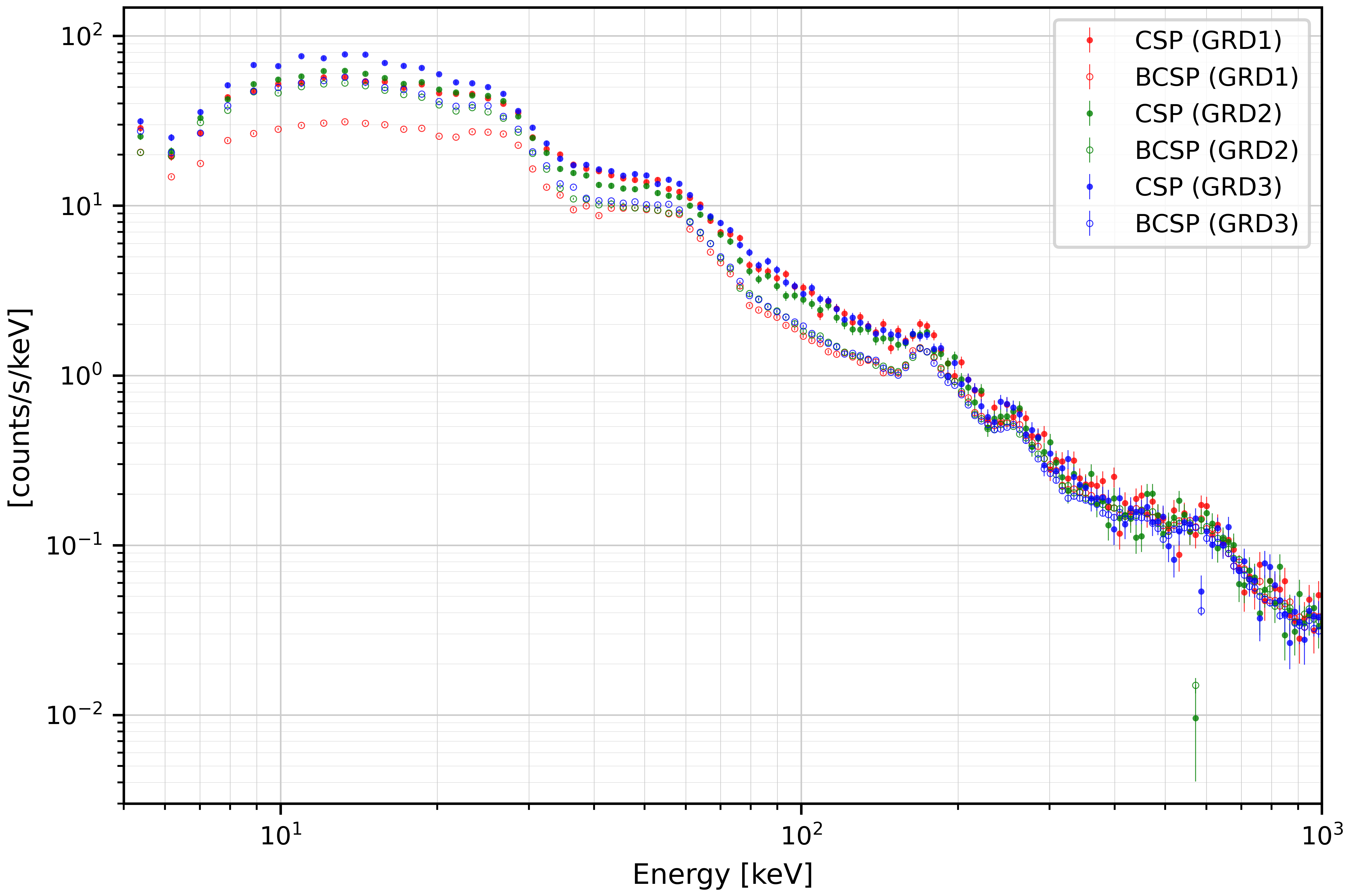}
  \caption{
    \label{fig:241018a-xband_grm_csp}
     GRB\,241018A (burst-id sb24101802) analysis by the \egx pipeline: time-integrated total count spectra (CSP,
     filled circles) and background count spectra (BCSP, open circles) in the three GRD detectors of the GRM, 
     for the time interval corresponding to the GRB $T_\mathrm{90}$ duration in ECLAIRs.
  }
\end{figure}
\begin{figure*}[t!]
  \centering
  \includegraphics[width=\linewidth]{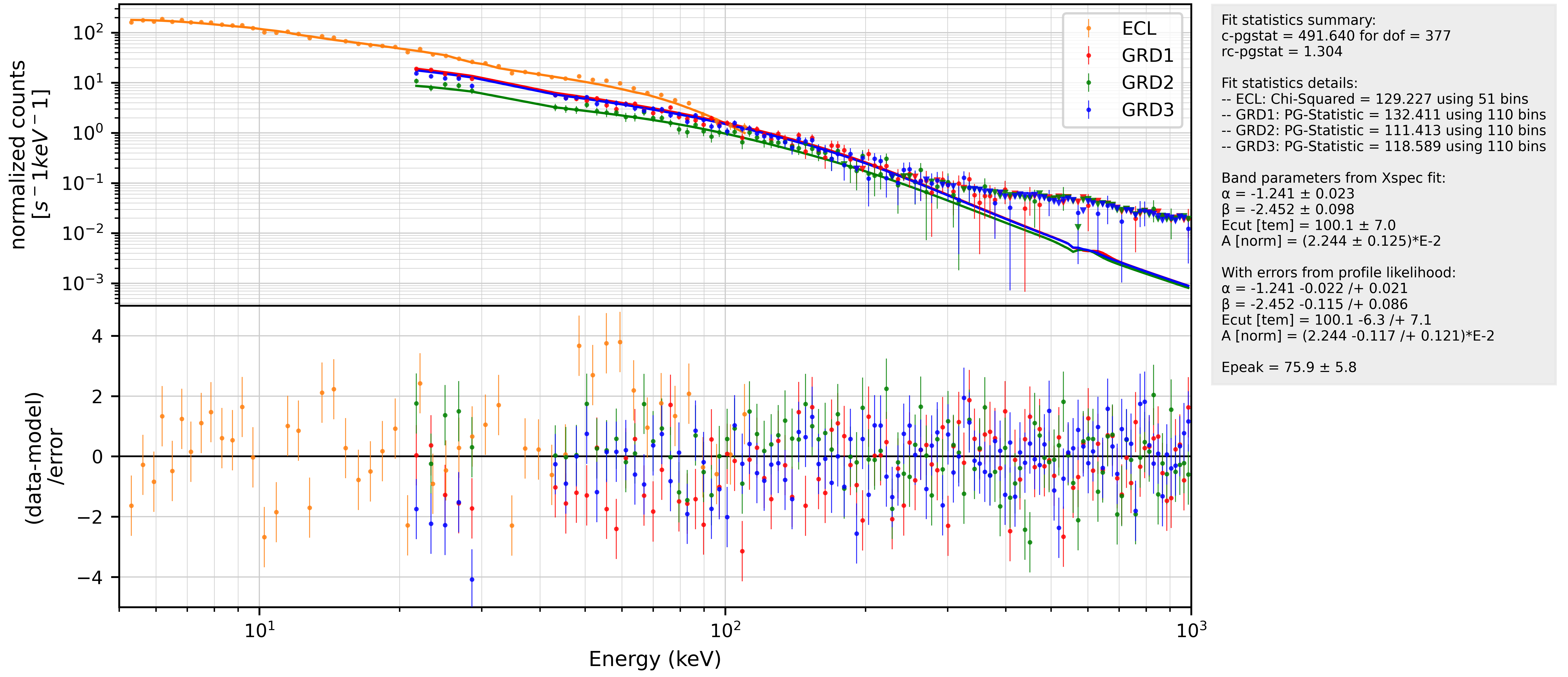}
  \caption{
    \label{fig:241018a-xband_fit}
     GRB\,241018A (burst-id sb24101802): time-integrated joint spectral fit of ECLAIRs and GRM data with the Band
     function from 5\,keV to 1\,MeV, for the time interval corresponding to the GRB $T_\mathrm{90}$ duration in ECLAIRs.
     The top panel displays the GRB count spectrum in ECLAIRs obtained with the ECPI/SPEX module \fpm{\citep{raa-ecpi}},
     and the background-subtracted count spectrum in each GRD of the GRM (from the CSP$-$BCSP difference).
     Triangles indicate 95\% upper limits for the energy channels lacking significant signal.
     The continuous curves show the count spectra that are predicted in the detectors by folding the best-fit Band model
     with their spectral responses. 
     The bottom panel shows the normalized residuals of the joint fit.
     The GRM \textit{pgstat} and ECLAIRs \textit{chi} statistics are combined in the fit as described in the legend, which
     also gives the best-fit parameters and their 68\% errors obtained either from the error matrix or from their profile likelihood.
  }
\end{figure*}

The XBAND-ECL target is activated if ECLAIRs triggered and if ECL-EVT-CAL event data are available in
the FSC SDB. If successful, then the XBAND-ECLGRM target is activated provided GRM-EVT-TRIG event data are
available as well.
The successive tasks described below are performed for each target (unless specified).

\subsection{\fpm{Temporal Analysis Tasks}}

The Merge (MRG) task is specific to the XBAND-ECL target.
It creates the missing \fpm{ECLAIRs calibrated event files (if any)} by running the EPCI software, and merges them into a single file
(ECL-EVT-CAL-MRG).
Then, it identifies the pixels and surface of the ECLAIRs detection plane that were illuminated by the
GRB before the satellite slew, where the source position is stable in the field of view.

The ECLAIRs Position (PO\_ECL) task first creates informative plots, like the QLC task (see section~\ref{sect:vhf}).
The task is called a second time after the T90 task (see below) to compute the GRB position through image
deconvolution by running the ECPI/IMAG module \fpm{\citep{raa-ecpi}} on the ECLAIRs $T_\mathrm{90}$ time interval.

The Count Light Curve (CLC) task bins ECLAIRs and GRM calibrated event data in time and in different energy bands.
Unlike VHF OBLCs, the time resolution of the X-band CLCs and their energy bands can be set freely. Moreover, the
ECLAIRs CLCs can be built by considering only the pixels in the detector plane which were illuminated by the GRB.
This method reduces the background noise while preserving most of the source signal.
In automated mode, the \egx pipeline uses the low/mid/best time resolutions defined in Table~\ref{tab:eclgrm-vhf}, and the
following energy bands (in keV): 4--20, 20--50, 50--80, 80--120, 4--120 for ECLAIRs; and 15--50, 50--300,
300--550, 550--5000, 15--550 for the GRM.
Once created, the CLCs are analyzed with the same algorithms as in the QLC task (see section~\ref{sect:vhf}), as
shown by the BSCLCs in Figure~\ref{fig:241018a-xband_eclgrm_clc}.

The T90 duration task estimates the GRB $T_\mathrm{90}$, $T_\mathrm{80}$ and $T_\mathrm{50}$ durations, as well as the on-ground
significance, of the GRB signal from the BSCCLCs of ECLAIRs and GRM, with the same algorithms as in the QT90
task (see section~\ref{sect:vhf}).
In automated mode, these quantities are computed in the 4--120\,keV and 50--300\,keV energy bands for ECLAIRs and
the GRM, respectively.
In addition, the T90 task defines the time intervals (episodes hereafter) for the time-integrated
(over the ECLAIRs $T_\mathrm{90}$ by default) and time-resolved spectral analyses.

\subsection{\fpm{Spectral Analysis Tasks}}
\begin{figure*}[t!]
  \centering
  \includegraphics[width=\linewidth]{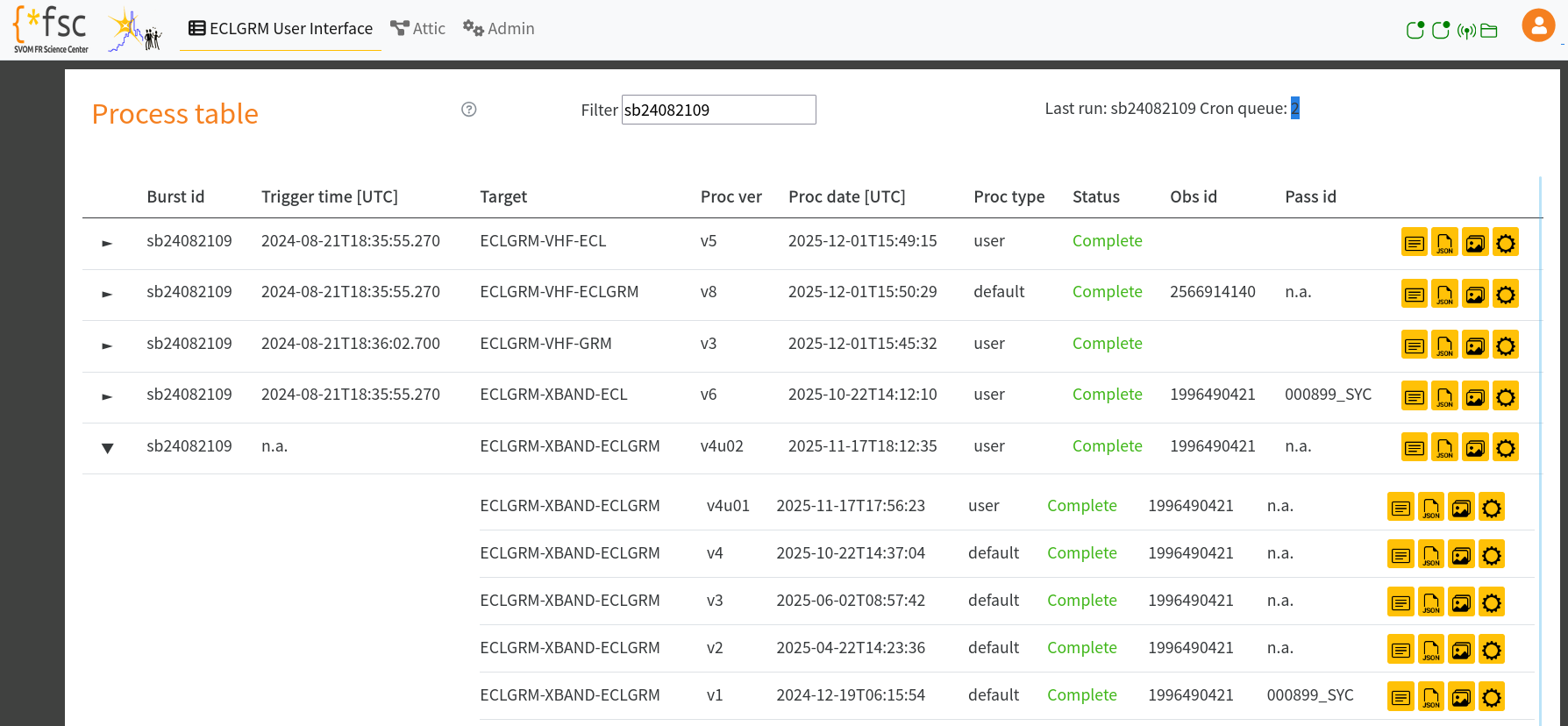}
  \caption{
    \label{fig:eclgrm-ui}
    Main page of the \egui user interface, for GRB\,240821A (burst-id sb24082109).
    Different target executions (processes) are displayed, with four buttons (right side) to access the log, the
    internal dictionnary, the dashboard and the analysis configuration editor (gear wheel).
    Processes marked as v[n]u[mm] correspond to the manual reprocessings where the output \fpm{SDP}s were not sent to the FSC
    SDB. 
  }
\end{figure*}

For each episode, the CSP, BCSP, ICSP, DRM and IDRM tasks compute the input \fpm{SDP}s of the spectral analysis.
Nearly logarithmic PI (pulse-invariant) energy channels are first defined, which follow the instrument's energy
resolution.
Then, two techniques are employed to build the count spectra: the counting technique, which is suitable for the GRM
and can also be used for ECLAIRs, and the imaging technique, which is preferred for ECLAIRs.
In the counting technique, calibrated event data
(ECL-EVT-CAL-MRG and GRM-EVT-TRIG) are binned to produce the total count spectrum (CSP task) in each detector.
The corresponding background count spectra (BCSP task) are obtained by creating a count light curve for each channel,
and fitting it in the background time intervals defined by the CLC task,
using a Poisson likelihood to account for possibly low count rates (see Figure~\ref{fig:241018a-xband_grm_csp}).
In the imaging technique, one single GRB count spectrum (ICSP task) is built by
running the ECPI/SPEX module \fpm{\citep{raa-ecpi}} on a selected range of energy channels.
Finally, the detector response matrices that are appropriate for each technique (DRM or IDRM tasks) are obtained from
the ECLAIRs CALDB or from the GRM DRM API, and rebinned to match the energy channels.

For each episode, the spectral (SP) analysis task uses the Python interface of the Xspec
software\footnote{\url{https://heasarc.gsfc.nasa.gov/docs/software/xspec/index.html}} to compute the GRB photon
spectrum (in $\mathrm{ph}\,\mathrm{cm}^{-2}\,\mathrm{s}^{-1}\,\mathrm{keV}^{-1}$), assuming different spectral
models.
While custom models will be available in manual mode, the automated mode considers only three models: the Band  
function \citep{Band+1993} and its asymptotic limits (power-law with or without exponential cutoff).
\fpm{The three spectral fits are performed using the proper statistics, namely \textit{pgstat} or \textit{chi}
  statistics\footnote{See Appendix B of the Xspec documentation:
  \url{https://heasarc.gsfc.nasa.gov/docs/software/xspec/manual/XspecManual.html}} 
  for count spectra obtained with the counting or imaging technique, respectively}.
Since these models are nested, the SP task then performs a likelihood ratio test to identify the best-fit
model.
Finally, the task computes the associated (photon and energy) fluxes and fluences in various energy bands.

Figure~\ref{fig:241018a-xband_fit} shows the time-integrated count spectra of GRB\,241018A in ECLAIRs and the GRM, jointly
fitted by the Band model from 5\,keV to 1\,MeV (SP task of the XBAND-ECLGRM target).
The high quality of the fit is indicated by the low value of the reduced fit statistics ($1.3$), and by the good residuals
across the whole spectrum, without adding any systematic uncertainties on the spectral responses despite the high
signal-to-noise ratio of the spectrum and the flux statistical accuracy of a few percent only.
\fpm{According to the SVOM mission high-level requirements, the ECLAIRs and GRM flux calibration should be precise to $\pm$20\%
  for transient sources (in their nominal energy range). The fit displayed in Figure~\ref{fig:241018a-xband_fit} is thus remarkable,
  since systematic errors that are much smaller than the requirements would suffice to flatten the residuals.}

\subsection{\fpm{Final Tasks}}
For each episode, the Hardness Ratio (HR) task computes the ratio between two energy bands (high/low) of the GRB signal 
measured in counts, photon fluence and energy fluence.
In automated mode, the HR task uses the following high/low energy bands (in keV):
[80--120]/[4--20], [80--120]/[20--50] and [80--120]/[50--80] for the XBAND-ECL target, and [50--300]/[4--120] for the
XBAND-ECLGRM target, respectively.

The LAG task computes the time lags between the count light curves (from the CLC task) in separate energy bands, following
the method developed by \citep{Bernardini+2015}.
For each pair of light curves, a modified cross-correlation \fpm{function (CCF, \cite{Band+1997})} is calculated for a
series of time delays, and the lag is defined as the time delay that maximizes the CCF.
The uncertainties on the CCF are derived by applying a flux-randomization method \citep{Peterson+1998}.
In automated mode, the LAG task uses the same high/low energy bands as the HR task.

\section{The User Interface for the ECLAIRs-GRM Instrument Scientist}
\label{sect:ui}

The \egui user interface lists all the processes executed in the \eg pipelines (see Figure~\ref{fig:eclgrm-ui}).
Each process corresponds to a given SVOM burst-id and to a given \eg target.
The burst-id is generated by the FSC VHF manager \citep{raa-fsc} for each new GRB candidate that triggered ECLAIRs
and/or the GRM.
In automated mode, each target can be executed several times, as complementary VHF or X-band data are received on
ground and the \eg input \fpm{SDP}s are updated in the SDB by the FSC pre-processing or ECPI pipelines.
The \egv and \egx pipelines run sequentially, queuing each new process until the current one completes.

The \egui user service has several functionalities. It gives access to the configuration file and log of each process, and
offers a dedicated dashboard for each of the five targets to visualize the analysis results and their associated control
plots.
Moreover, the \egis can restart part or all of the analysis tasks interactively through the \egui
configuration editor, to optimize the results with a different choice of parameters.
In this manual mode, the output \fpm{SDP}s are sent to the FSC SDB only if desired, which automatically shares them with the
SVOM Burst Advocate through the iFSC-Tools (see Figures~\ref{fig:eclgrm-vhf} and~\ref{fig:eclgrm-xband}).
The \egis is responsible for the validation of the \eg \fpm{SDP}s, for their report in \fpm{GCN (General
  Coordinates Network)} circulars and for their display in the SVOM GRB public table \citep{raa-fsc-tools}.

\section{Conclusions}
\label{sect:conclusion}
The two \eg data processing pipelines have been running at the FSC since the launch of the SVOM satellite on June 22,
2024.
They produce the (real-time) VHF and (complete) X-band \fpm{scientific data products} (light curves, durations, spectra, etc.) related to
the GRB prompt high-energy emission (from 4\,keV to 5\,MeV), combining the ECLAIRs and GRM observations in joint analyses.
For each GRB candidate detected by both instruments, the \eg pipelines generate more than forty scientific \fpm{data}
products and store them in FITS format in the FSC SDB.

The \egui user interface has been also used \fpm{extensively to} monitor the data processings in the
pipelines and to refine the results whenever needed.
The three \eg services thus allow the \egis to support SVOM Burst Advocate activity in real time, to
contribute to the GCN circulars on ECLAIRs and GRM, and to provide scientifically-validated products for the SVOM GRB
public table.

While the main functionalities of the pipelines are in place, and their performance demonstrated on past GRB events,
more developments are envisaged to complement the existing products.
Not exhaustively, this includes computing the emission durations in additional energy bands, improving the classification
algorithm, developing an automated identification of the emission episodes for the time-resolved spectral analysis,
and accounting for the systematic uncertainties on the spectral responses \fpm{(as discussed in \cite{raa-ecl-inst})} in
the reconstruction of GRB spectra.

\begin{acknowledgements}
The Space-based multi-band astronomical Variable Objects Monitor (SVOM) is a joint Chinese-French mission led by the
Chinese National Space Administration (CNSA), the French Space Agency (CNES), and the Chinese Academy of Sciences
(CAS). We gratefully acknowledge the unwavering support of NSSC, IAMCAS, XIOPM, NAOC, IHEP, CNES, CEA, and CNRS. 
\end{acknowledgements}

\appendix                  
\section{\fpm{List of Acronyms}}
\label{sect:appendix}
\begin{table}[h!]
  \begin{center}
    \caption{
      \fpm{List of Acronyms and their Definition.}
    }
    \begin{tabular}{ll}
      \hline\noalign{\smallskip}
      Acronym& Definition\\ 
      \hline\noalign{\smallskip}
      API & Application programming interface\\
      CALDB & Calibration data base\\
      BSCLC & Background-subtracted count light curve\\
      BSCCLC & Background-subtracted cumulative count light curve\\
      BCSP & Background count spectrum (counting technique)\\
      CCF & Cross-correlation function\\
      CLC & Count light curve\\
      CSP & Total count spectrum (counting technique)\\
      CRCLASS & Crude classification\\
      DRM & Detector response matrix (counting technique)\\
      ECL-EVT-CAL & ECLAIRs calibrated event data\\
      ECLGRM & ECLAIRs-GRM system\\
      ECLGRM-IS & ECLGRM instrument scientist\\
      ECLGRM-VHF & ECLGRM quick analysis pipeline (VHF data)\\
      ECLGRM-XBAND & ECLGRM complete analysis pipeline (X-band data)\\
      ECLGRM-UI & ECLGRM user interface\\
      ECPI & ECLAIRs pipeline\\
      FSC & French science center\\
      GCN & General coordinates network\\
      GRM & Gamma-ray monitor\\
      GRM-EVT-TRIG & GRM calibrated event data\\
      GRM-IC & GRM instrument center\\
      GRD & Gamma-ray detector\\
      HR & Hardness ratio\\
      ICSP & Source count spectrum (imaging technique)\\
      IDRM & Detector response matrix (imaging technique)\\
      IS & Instrument scientist\\
      MRG & Merge\\
      OBLC & Onboard light curve\\
      QHR & Quick hardness ratio\\
      QLC & Quick light curve\\
      QPF & Quick peak flux\\
      SDB & Science data base\\
      SDP & Scientific data product\\
      SP & Spectrum\\
      VHF-ECL & ECLAIRs target of the \egv pipeline\\
      VHF-ECLGRM & ECLGRM joint target of the \egv pipeline\\
      VHF-GRM & GRM target of the \egv pipeline\\
      XBAND-ECL & ECLAIRs target of the \egx pipeline\\
      XBAND-ECLGRM & ECLGRM joint target of the \egx pipeline\\
      \hline\noalign{\smallskip}
    \end{tabular}
  \end{center}
\end{table}


\label{lastpage}


\begin{thebibliography}{99}

\bibitem[Band et al.(1993)]{Band+1993} Band D. L. et al., 1993, \apj, 413, 281

\bibitem[Band et al.(1997)]{Band+1997} Band D. L. et al., 1997, \apj, 486, 928
  
\bibitem[Bernardini et al.(2015)]{Bernardini+2015} Bernardini M. G. et al. 2015, \mnras, 446, 1129

\bibitem[Bernardini et al.(2017)]{Bernardini+2017} Bernardini M. G. et al. 2017, Experimental Astronomy, 44, 113

\bibitem[Claret et al.(2025)]{raa-fsc-tools} Claret A. et al., RAA 2025, 25, this issue

\bibitem[Daigne et al.(2025)]{raa-grbs} Daigne F. et al., RAA 2025, 25, this issue
  
\bibitem[Godet et al.(2025)]{raa-ecl-inst} Godet O. et al., RAA 2025, 25, this issue

\bibitem[Dong et al.(2025)]{raa-mission-satellite} Dong L. et al., RAA 2025, 25, this issue

\bibitem[Cordier et al.(2025a)]{raa-mission-space} Cordier B. et al., RAA 2025a, 25, this issue

\bibitem[Cordier et al.(2025b)]{raa-vhf-system} Cordier B. et al., RAA 2025b, 25, this issue

\bibitem[Goldwurm et al.(2025)]{raa-ecpi} Goldwurm A. et al., RAA 2025, 25, this issue

\bibitem[He et al.(2025)]{raa-grm-trigger} He J. et al., RAA 2025, 25, this issue

\bibitem[Louvin et al.(2025)]{raa-fsc} Louvin H. et al., RAA 2025, 25, this issue

\bibitem[Maïolo(2023)]{Maiolo-phd} Maïolo A. 2023, PhD thesis

\bibitem[Mate et al.(2019)]{Mate+2019} Mate S. et al. 2019, Experimental Astronomy, 48, 171

\bibitem[Oganesyan et al.(2017)]{Oganesyan+2017} Oganesyan G. et al., 2017, \apj, 846, 137

\bibitem[Peterson et al.(1998)]{Peterson+1998} Peterson B. M. et al. 1998, \pasp, 110, 660
  
\bibitem[Scargle et al.(2013)]{Scargle+2013} Scargle J. D. et al., 2013, \apj, 764.2, 167

\bibitem[Schanne et al.(2025)]{raa-ecl-trigger} Schanne S. et al., RAA 2025, 25, this issue

\bibitem[Sun et al.(2025)]{raa-grm-inst} Sun J.-C. et al., RAA 2025, 25, this issue

\bibitem[Vianello (2018)]{Vianello2018} Vianello G. 2018, \apjs, 236 17

\bibitem[Wang et al.(2025)]{raa-grm-pipeline} Wang P. et al., RAA 2025, 25, this issue

\bibitem[Wei et al.(2016)]{white-paper} Wei J., Cordier B. et al. 2016, arXiv:1610.06892

\bibitem[Wei et al.(2025)]{raa-mission-ground} Wei J. et al., RAA 2025, 25, this issue  
  
\bibitem[Yurong et al.(2025)]{raa-ground-system} Yurong L. et al., RAA 2025, 25, this issue

\bibitem[Zhao et al.(2025)]{raa-grm-rsp} Zhao X.-Y. et al., RAA 2025, 25, this issue
 
\end{thebibliography}
\end{document}